\documentclass[hyper]{JHEP} 

\usepackage{epsfig}

\newcommand\fverb{\setbox\pippobox=\hbox\bgroup\verb}

\newcommand\fverbdo{\egroup\medskip\noindent%

            \fbox{\unhbox\pippobox}\ }

\newcommand\fverbit{\egroup\item[\fbox{\unhbox\pippobox}]}

\newbox\pippobox

\title{Branes at Quantum Criticality}
\author{J. Kluso\v{n}\\
Department of
Theoretical Physics and Astrophysics\\
Faculty of Science, Masaryk University\\
Kotl\'{a}\v{r}sk\'{a} 2, 611 37, Brno\\
Czech Republic\\
E-mail: \email{klu@physics.muni.cz}}
\preprint{\hepth{0904.1343}}

 \abstract{In this paper we propose     new non-relativistic
$p+1$ dimensional  theory. This theory is defined
in such a way that the potential term
obeys the principle of detailed balance where the
generating action corresponds  to
 p-brane action. This condition ensures
that the norm of the vacuum wave functional of $p+1$ dimensional
theory is equal to the partition function of
p-brane theory.} 

\def\mH{\mathcal{H}}
\def \Gi{\left(G^{-1}\right)}

\def\bx{\mathbf{x}}
\def\by{\mathbf{y}}

\def\ket #1{\left|#1\right>}

\newcommand{\bT}{\mathbf{T}}

\newcommand{\mL}{\mathcal{L}}

\def\pb #1{\left\{#1\right\}}
\begin{document}
\section{Introduction}\label{first}
Recently P. Ho\v{r}ava proposed a  new
intriguing formulation of theories with
anisotropic scaling between time and
spatial dimensions
\cite{Horava:2009if,Horava:2009uw,Horava:2008ih,Horava:2008jf}
\footnote{For recent study of
cosmological aspects of these theories,
see
\cite{Kiritsis:2009sh,Calcagni:2009ar,Takahashi:2009wc}.}.
In particular, in  his second paper
\cite{Horava:2008ih} he formulated a
new world-volume quantum theory of
gravity and matter in $2+1$ dimensions
that is strongly anisotropic between
space and time in the world-volume
theory. This phenomenon is well known
from the study of condensed matter
systems at quantum criticality
\cite{Ardonne:2003wa}. Similar systems
have  been intensively studied from the
point of view of the nonrelativistic
form of the AdS/CFT correspondence
\cite{Son:2008ye,Balasubramanian:2008dm,Goldberger:2008vg,
Barbon:2008bg,Wen:2008hi,Herzog:2008wg,Maldacena:2008wh,
Adams:2008wt,Minic:2008xa,Chen:2008ad,Colgain:2009wm,Bagchi:2009my,
Alishahiha:2009hg,Donos:2009en,Pal:2009yp,Danielsson:2008gi,
Taylor:2008tg,Adams:2008zk,Akhavan:2008ep,Rangamani:2008gi,Schvellinger:2008bf,
Sachdev:2008ba,Hartnoll:2008rs,Lin:2008pi,Yamada:2008if,Duval:2008jg,
Kovtun:2008qy,Kachru:2008yh}
 \footnote{For another approach to the
 study of
non-relativistic systems in string
theory,see for example
\cite{Lee:2009mm,Galajinsky:2009dw,Nakayama:2009cz,
Nakayama:2008qm,Gomis:2004pw,Brugues:2004an,Kluson:2006xi,
Gomis:2006wu,Gomis:2006xw,Sakaguchi:2006pg,Gomis:2005bj,Gomis:2005pg}.}.
The construction of such a 2+1
dimensional theory \cite{Horava:2008ih}
was based on the following question: Is
it possible to find a quantum theory of
membranes such that its ground state
wave functional reproduces the
partition function of the bosonic
string? Generally, we can start with
some equilibrium system in $D$
dimensions that is at criticality and
study  how the critical behavior
extends to the dynamical phenomena in
$D+1$ dimensions.  For example, the
similar question can be asked in the
context of stochastic quantization; the
goal is to build a non-equilibrium
system in $D+1$ dimensions that relaxes
at late times to its ground state,
which reproduces the partition function
of a $D$ dimensional system we are
interested in.

The goal of this paper is to implement
 similar ideas in the case where the
$p$ dimensional system is a brane with
the  Nambu-Goto form of the action. We
also assume that this p-brane is
embedded in general $D$ dimensional
background. This assumption implies
that the $p$ dimensional action is
highly non-linear with all well known
consequences for the renormazibility
and quantum analysis of given action.
Despite of this fact we
 demand-with analogy with the quantum critical
membrane theory
\cite{Horava:2008ih}-that the partition
function of $p$ dimensional theory
should be equivalent to the norm of the
ground state of the  $p+1$ dimensional
theory. Note however that this
correspondence is pure formal since
 we do not address the question whether these
objects for highly nonlinear systems
are really well defined. Despite of
this fact we proceed further and we
will see that the fundamental
requirement allows us to find an action
for the $p+1$ dimensional theory that
is manifestly invariant under
\emph{spatial diffeomorphism} and under
\emph{rigid time translation}. Further,
the resulting action obeys the
\emph{Detailed balance condition}  that
states that the potential term of $p+1$
dimensional theory can be derived from
the variation principle of $p$
dimensional theory.

As the next step in the construction of
the action for a p-brane at criticality
we  extend the symmetries of given
action. More precisely, we extend the
time-independent spatial
diffeomorphisms to all space-time
diffeomorphisms that respect the
preferred codimension-one foliation of
$p+1$ dimensional space-time by the
slices at fixed time. These
diffeomorphism are known as a
foliation-preserving diffeomorphism and
consist a space-time dependent spatial
diffeomorphism together with
time-dependent reparameterization of
time. Since under these transformations
the original $p+1$ dimensional
 action is not invariant
we have to introduce new gauge fields $N^i$
and $N$ to maintain its invariance.
  The presence of these new gauge
fields will be crucial for the correct
Hamiltonian formulation of the theory
as the theory of constraint systems. In
fact, since the action does not contain
time derivative of $N$ and $N^i$ the
standard analysis implies an existence
of primary constraints $\pi_N\approx 0
\ , \pi_i\approx 0$. The consistency of
these constrains implies an existence
of secondary constraints.  Then we
construct vacuum wave functional that
 is annihilated by these  constraints and that
 is automatically the state of zero energy.

The organization of this paper is as
follows. In the next section
(\ref{second}) we review the  Lifshitz
theory of $D$ scalar fields defined on
$p$ dimensional space. In section
(\ref{third}) we construct the $p+1$
dimensional theory from the Nambu-Goto
form of p-brane action that obeys
\emph{detailed balance condition}. In
section (\ref{fourth}) we generalize
the gauge symmetries of this
$p+1$-dimensional theory when we extend
rigid time translation and spatial
diffeomorphism to the
foliation-preserving diffeomorphism. In
section (\ref{fifth}) we develop the
Hamiltonian formalism for given theory
and we calculate the algebra of
constraints. Finally in section
(\ref{sixth}) we outline our results
and suggest possible extension of this
work.
\section{Review of Lifshitz Scalars and
Quantum Criticality} \label{second} The
aim of this section is to review,
following
\cite{Horava:2008ih,Ardonne:2003wa} the
physics  of $D$ free scalar fields
defined on $p$ dimensional Euclidean
space with coordinates $\bx=x^i \ ,
i=1,\dots,p$ with   action
\begin{equation}
W=\frac{1}{2}\int d^p\bx \delta^{ij}
\partial_i\Phi^M\partial_j \Phi^N
g_{MN} \ ,
\end{equation}
where  $g_{MN}$ is a \emph{constant
positive definite symmetric} matrix.

As in standard quantum mechanics, the
fundamental object of this
 theory is the
partition function $\mathcal{Z}$
\begin{equation}\label{Zi}
\mathcal{Z}=\int \mathcal{D} \Phi(\bx)
\exp[-W(\Phi(\bx))] \ ,
\end{equation}
that is defined as
 a path integral on the space of
field configurations $\Phi^M(\bx)$. Let
us  assume the existence of a $p+1$
dimensional theory whose configuration
space coincides with the space of all
$\Phi^M(\bx)$. In other words, the wave
functionals of this $p+1$-dimensional
theory are functionals of $\Phi^M(\bx)$
so that $\Psi(\Phi^M(\bx))$.
 Then the
standard interpretation of quantum
mechanics implies that $\Psi(\Phi(\bx))
\Psi^*(\Phi(\bx))$ is a density on the
configuration space. Our goal is to
formulate $p+1$ dimensional system with
the property that the norm of its
ground-state functional
$\Psi_0(\Phi(\bx))$ reproduces the
partition function (\ref{Zi})
\begin{equation}
\left<\Psi_0|\Psi_0\right>= \int
\mathcal{D}\Phi(\bx)
\Psi_0^*(\Phi(\bx)) \Psi_0(\Phi(\bx))
=\int \mathcal{D} \Phi(\bx)
\exp[-W(\Phi(\bx))] \ .
\end{equation}
To proceed further we have to introduce
 of Schr\"{o}dinger formulation
of quantum field theory. Explicitly,
the
basic operators in quantum
field theory  are  $\hat{\Phi}^M(\bx)$
and $\hat{\Pi}_M(\bx)$ with canonical
commutation relation
\begin{equation}\label{com}
[\hat{\Phi}^M(\bx),\hat{\Pi}_N(\by)]=
i\delta^M_N\delta(\bx-\by) \ .
\end{equation}
Further, the eigenstates of
$\hat{\Phi}^M(\bx)$ are the states
$\ket{\Phi(\bx)}$ that obeys
\begin{equation}
\hat{\Phi}^M(\bx)\ket{\Phi(\bx)}=
\Phi^M(\bx)\ket{\Phi(\bx)} \ .
\end{equation}
In the Schr\"{o}dinger  representation
any state of given system is
represented as the state functional
$\Psi(\Phi(\bx))$ and the action of the
operator $\hat{\Phi}^M(\bx)$ on this
state functional corresponds to
multiplication with $\Phi^M(\bx)$.
Further, the commutation relation
(\ref{com}) implies that in the
Schr\"{o}dinger representation the
operator $\hat{\Pi}_M(\bx)$ is equal to
\begin{equation}
\hat{\Pi}_M(\bx)=-i\frac{\delta}{\delta
\Phi^M(\bx)} \ .
\end{equation}
Let us now assume that the Hamiltonian
of the $p+1$ dimensional theory has the
form
\begin{equation}\label{defH}
\hat{H}=\frac{1}{2}\int d^p\bx \hat{\mH}\equiv \int
d^p\bx \hat{Q}_M^\dag(\bx) \hat{g}^{MN}\hat{Q}_N(\bx) \ ,
\end{equation}
where  $\hat{Q}_M,\hat{Q}_M^\dag$ are defined as
\begin{equation}\label{defQ}
\hat{Q}_M=i\hat{\Pi}_M+\frac{1}{2}\frac{\delta
W[\hat{\Phi}]}{\delta
\hat{\Phi}^M(\bx)} \ , \quad \hat{Q}_M^\dag=
-i\hat{\Pi}_M+\frac{1}{2}\frac{\delta
W[\hat{\Phi}]}{\delta
\hat{\Phi}^M(\bx)} \ .
\end{equation}
Clearly, the Hamiltonian (\ref{defH})
is  Hermitian and positive. Note also
that $\hat{Q}_M,\hat{Q}_M^\dag$  take
the form in  the Schr\"{o}dinger
representation
\begin{equation}\label{QMshch}
\hat{Q}_M=\frac{\delta}{\delta\Phi^M(\bx)}+\frac{1}{2}\frac{\delta
W[\Phi]}{\delta \Phi^M(\bx)} \ , \quad
\hat{Q}_M^\dag=
-\frac{\delta}{\delta\Phi^M(\bx)}+\frac{1}{2}\frac{\delta
W[\Phi]}{\delta \Phi^M(\bx)} \ .
\end{equation}
Let us assume that   the vacuum wave
functional takes the form
\begin{equation}
\Psi_0(\Phi(\bx))= \exp(-\frac{1}{2}W)
=\exp(-\frac{1}{4}\int d^p\bx
\delta^{ij}\partial_i\Phi^M(\bx)g_{MN}\partial_j
\Phi^N(\bx)) \ .
\end{equation}
It is easy to see that $\hat{Q}_M$
defined in (\ref{QMshch}) annihilates
$\Psi_0$
\begin{equation}
\hat{Q}_M\Psi(\Phi(\bx))=0
\end{equation}
as follows from the fact that
\begin{equation}
\frac{\delta }{\delta \Phi^M(\bx)}
\Psi_0(\Phi)=-\frac{1}{2}\frac{\delta
W}{\delta \Phi^M(\bx)}\Psi_0(\Phi) \ .
\end{equation}
Using the definition of the Hamiltonian
(\ref{defH}) we derive that the vacuum
state functional is the
 eigenstate of the Hamiltonian
of zero energy.

As the next step in the construction
of the $p+1$-dimensional theory we
should find corresponding Lagrangian
density from the known  quantum
Hamiltonian (\ref{defH}).
The standard procedure is
to consider the classical form of this
Hamiltonian when we identify
$\hat{\Pi}\rightarrow \Pi$ and
$\hat{\Phi}\rightarrow \Phi$ so
that the classical Hamiltonian
is equal to
\begin{eqnarray}
H&=&\frac{1}{2}\int d^p\bx
\left(-i\Pi_M(\bx)+\frac{1}{2}\frac{\delta
W}{\delta\Phi_M(\bx)})g^{MN}
(i\Pi_M(\bx)+ \frac{1}{2}\frac{\delta
W}{\delta\Phi_M(\bx)}\right) =\nonumber \\
&=&\frac{1}{2}\int d^p\bx \left(\Pi_M(\bx)
g^{MN}\Pi_N(\bx)+\frac{1}{4}
\partial_i[\delta^{ij}g_{MN}\partial_j\Phi^N(\bx)]g^{MK}
\partial_i[\delta^{ij}g_{KL}\partial_j\Phi^L(\bx)]
\right)  \ ,
\nonumber \\
\end{eqnarray}
where we used the explicit form of the
variation
\begin{equation}
\frac{\delta W}{\delta \Phi^M(\bx)}=
-\partial_i[\delta^{ij}g_{MN}\partial_j\Phi^N(\bx)]
\ .
\end{equation}
Then in order to find corresponding Lagrangian
we use the Hamiltonian equation of motion
\begin{equation}
\partial_t\Phi^M(\bx)=
\pb{\Phi^M(\bx),H}=g^{MN}\Pi_N(\bx)
\end{equation}
and consequently
\begin{eqnarray}\label{Lagfr}
\mL &=&\partial_t\Phi^M\Pi_M-\mH=\nonumber \\
&=& \frac{1}{2}
\partial_t \Phi^M g_{MN}\partial_t \Phi^N
-\frac{1}{2}\left(\frac{1}{2}
\frac{\delta W}{\delta\Phi^M}\right)
g^{MN} \left(\frac{1}{2} \frac{\delta
W}{\delta \Phi^N}\right) =\nonumber \\
&=&\frac{1}{2}
\partial_t \Phi^M g_{MN}\partial_t \Phi^N-
\frac{1}{8}
\partial_i[\delta^{ij}g_{MN}\partial_j\Phi^N]g^{MK}
\partial_i[\delta^{ij}g_{KL}\partial_j\Phi^L]
\ .
\nonumber \\
\end{eqnarray}
We derived exactly as in
\cite{Horava:2008ih,Ardonne:2003wa}
that the  Lagrangian density of the
$p+1$ dimensional theory is a sum of a
kinetic term that involves time
derivative of $\Phi^M$ and a potential
term that is derived from the
variational principle.
This important property
is known as   \emph{detailed balance
condition}.

The theory defined by the Lagrangian
density (\ref{Lagfr}) has many
interesting properties. Firstly, if we
define the scaling dimension of $\bx$
as
\begin{equation}
[\bx]=-1
\end{equation}
we find from the requirement of the
invariance of $W$ under scaling that
the scaling dimension of $\Phi$ is
\begin{equation}
[\Phi]=\frac{p-2}{2} \ .
\end{equation}
However then from the requirement that
the scaling dimension of $p+1$
dimensional action is zero implies that
the scaling dimension of $t$ is
\begin{equation}
[t]=-2 \ .
\end{equation}
It is known from the theory of
condensed matter systems that the
degree of anisotropy between the time
and space is measured by the dynamical
critical exponent $z$ in the sense that
$[t]=-z$. Lorentz symmetry of
relativistic systems imply $z=1$ while
for non-relativistic systems we have
$z=2$.
\section{Branes at Criticality}\label{third}
In this section we generalize the analysis
presented in the previous section to the
case when  $p$ dimensional
theory is p-brane  with
 Nambu-Goto  action
\begin{equation}\label{Wact}
W=\frac{1}{\kappa^p_W} \int d^p\bx
\sqrt{\det G} \ , G_{ij}= g_{MN}
\partial_i \Phi^M\partial_j \Phi^N \ ,
\end{equation}
where the world-volume is labeled by
coordinates $x^i \ , i,j=1,\dots,p$.
Further, $\Phi^M \ , M=1,\dots,D$ are
 scalar fields (from the point
of view of  dimensional
world-volume theory) and
$g_{MN}(\Phi)$ is general
metric of target $D-$dimensional space.

Our goal is to find $p+1$-dimensional
theory  with the property that the
potential term obeys \emph{detailed
balance condition}. We proceed in the
similar way as in previous section
 and   demand that when we quantize this
theory on $R^{p,1}$ the resulting vacuum wave
functional should be equal to
\begin{equation}\label{Vacfun}
\Psi_0[\Phi(\bx)]=\exp\left(
-\frac{1}{2}W[\Phi(\bx)]\right) \ ,
\end{equation}
where $W$ is given in (\ref{Wact}). In
other words the wave functional is
function of $\Phi^M(\bx)$ that should
be defined on the space of gauge orbits
$\mathcal{A}/\mathcal{G}$ where
$\mathcal{A}$ is a space of all fields
$\Phi^M(\bx)$ and the gauge group
$\mathcal{G}$ is the group of
world-volume diffeomorphism.  Recall
that  the norm of the vacuum wave
functional (\ref{Vacfun}) is equal to
\begin{equation}\label{norm}
\int_{\mathcal{A}/\mathcal{G}}
\mathcal{D}\Phi[\bx] \Psi^*[\Phi(\bx)]
\Psi[\Phi(\bx)]=
\int_{\mathcal{A}/\mathcal{G}}
\mathcal{D}\Phi(\bx)
\exp\left(-W(\Phi(\bx))\right) \ .
\end{equation}
We see that
 this norm
  is   equal to the partition
function of  p-brane theory
 that  takes
the form
\begin{equation}\label{Z}
Z=\int_{\mathcal{A}/\mathcal{G}}
\mathcal{D}\Phi(\bx)
\exp\left(-W(\Phi(\bx))\right) \ .
\end{equation}
We should again stress
that (\ref{norm}) and  (\ref{Z})
are  formal prescriptions since
we did not precisely defined the
integration measure and the space
$\mathcal{A}/\mathcal{G}$.

Despite of these facts
 we  propose a
Hamiltonian of $p+1$ dimensional
 theory that has the property
 that the vacuum wave
functional (\ref{Vacfun})
 is its  ground state
with zero energy. Following
the analysis presented in
previous section we introduce operators
$\hat{\Phi}^M(\bx),\hat{\Pi}_M(\bx)$ that obey
the commutation relations (\ref{com})
and  define operators
$\hat{Q}_M(\bx),\hat{Q}_M^\dag(\bx)$ as
in (\ref{defQ}) where now  $W$ is given in
(\ref{Wact}). However due to the non-linear
character of (\ref{Wact}) it is clear that
the commutator of $\hat{Q}^\dag_M(\bx)$
 with $\hat{Q}_N(\by)$ is nonzero and
it is equal to some function
of $\hat{\Phi}^M$ together with
their spatial derivatives.
As a consequence  we have to  define
the quantum Hamiltonian with
the prescriptions that all
$\hat{Q}^\dag_M$'s are to the left of
all $\hat{Q}_N'$.
Explicitly, we propose the
quantum Hamiltonian in the form
\begin{eqnarray}\label{Hqm}
\hat{H}&=&\int d^p\bx \hat{\mH}(\bx) \ , \quad
\hat{\mH}=\frac{\kappa^2}{2}
\hat{Q}^\dag_M \frac{g^{MN}(\hat{\Phi})}
{\sqrt{\det G(\hat{\Phi})}}
\hat{Q}_N=\nonumber \\
&=&\frac{\kappa^2}{2}
\left(-i\hat{\Pi}_M+\frac{\delta W}{2\delta
\hat{\Phi^M}}\right)\frac{g^{MN}(\hat{\Phi})}
{\sqrt{\det
G(\hat{\Phi})}}
\left(i\hat{\Pi}_N+\frac{\delta W}{2\delta
\hat{\Phi}^N}\right) \ ,  \nonumber \\
\end{eqnarray}
where $\kappa$ is a coupling constant.
It is easy to see that the Hamiltonian
(\ref{Hqm}) is Hermitian and positive
definite. Further, it is also clear
that $\hat{Q}_M$ annihilates
$\Psi_0[\Phi(\bx)]$ given in
(\ref{Vacfun}) and consequently
(\ref{Vacfun}) is a candidate for the
ground state of the theory since by
definition
\begin{equation}
\hat{H}\Psi_0[\Phi(\bx)]=0 \ .
\end{equation}
Now we would like to find corresponding
Lagrangian formulation of given theory
defined by quantum mechanical
 Hamiltonian (\ref{Hqm}).
As in previous section we  consider the
classical version of this Hamiltonian
where we replace $\hat{\Pi}$ with $\Pi$
and $\hat{\Phi}$ with $\Phi$.  It is
clear that in this process we ignore
the
 ambiguity in the
ordering of $\Pi$ and $\Phi$ in the
Hamiltonian (\ref{Hqm}).
 With this
issue in mind  we claim that
the  classical form of  the Hamiltonian
density (\ref{Hqm}) takes the form
\begin{equation}\label{Hdcl}
\mH=\frac{\kappa^2}{2} \Pi_M
\frac{g^{MN}}{\sqrt{\det G}}
\Pi_N+\frac{\kappa^2}{8\sqrt{\det G}}
\frac{\delta W}{\delta \Phi^M}
g^{MN}\frac{\delta W}{\delta \Phi^N}
\ .
\end{equation}
Using this form of the Hamiltonian
density it is easy to determine the
Lagrangian density. Firstly
we determine the time derivative
of $\Phi^M$ from
\begin{equation}
\partial_t \Phi^M(\bx)=
\pb{\Phi^M(\bx),H} =\kappa^2
\frac{g^{MN}\Pi_N}{\sqrt{\det G}}
\end{equation}
and then we easily obtain the Lagrangian
density in the form
\begin{eqnarray}\label{Lkp}
\mL&=&\partial_\tau \Phi^M\Pi_M-\mH=
\mL_K-\mL_V \ , \nonumber \\
\mL_K&=&\frac{1}{2\kappa^2} \sqrt{\det
G}\partial_\tau \Phi^M g_{MN}
\partial_\tau \Phi^N \ , \nonumber \\
\mL_V&=&\frac{\kappa^2}{8\kappa_W^{2p}}\sqrt{\det
G} \left(\partial_M
G_{ij}G^{ji}-\frac{1}{\sqrt{\det G}}
\partial_i [g_{MK}\partial_j \Phi^K
G^{ji}\sqrt{\det G}]\right)g^{MN}\times
\nonumber \\
&\times &\left(\partial_N
G_{kl}G^{lk}-\frac{1}{\sqrt{\det G}}
\partial_k[g_{NL}\partial_l \Phi^L
G^{lk}\sqrt{\det G}]\right) \ .
\nonumber \\
\end{eqnarray}
Now we analyze the Lagrangian
density (\ref{Lkp}) in more details.
Let us consider the diffeomorphism
transformations
\begin{equation}\label{spdif}
x'^i=x'^i(\bx) \
\end{equation}
under which the  element
$d^p\bx$ transforms as
\begin{equation}
d^p\bx'=d^p\bx \left|\det D\right| \ ,
\end{equation}
where we introduced $p\times p$ matrix
$D^i_j=\frac{\partial x'^i}{\partial
x^j}$. Further, by definition
$\Phi^M(\bx)$ are world-volume scalars
so that they transform under
(\ref{spdif}) as
\begin{equation}
\Phi'^M(\bx')=\Phi^M(\bx) \ .
\end{equation}
It is easy to see that $G_{ij}$
transform in the following way
\begin{eqnarray}
 G'_{ij}(\Phi'(\bx'))&=&
G_{kl}(\Phi(\bx))(D^{-1})^k_i
(D^{-1})^l_j
\nonumber \\
\sqrt{\det G'(\Phi'(\bx') )}&=&
\frac{1}{|\det D|}\sqrt{\det
G(\Phi(\bx))}
\nonumber \\
\end{eqnarray}
and consequently the Lagrangian
densities $\mL_K,\mL_V$ transform as
\begin{equation}
\mL_K(\Phi'(\bx'))=\frac{1}{|\det
D(\bx)|}\mL_K(\Phi(\bx)) \ , \quad
\mL_V(\Phi'(\bx'))=\frac{1}{|\det
D(\bx)|}\mL_V(\Phi(\bx)) \ . \quad
\end{equation}
Using these results we immediately
obtain that  $p+1$ dimensional action
\begin{equation}\label{S}
S=\int d^p\bx dt \mL
\end{equation}
 is invariant under
spatial diffeomorphisms (\ref{spdif}).
We could also proceed in opposite
direction and   demand that the $p+1$
dimensional action should  be invariant
under (\ref{spdif}). However this
requirement implies that the expression
$\frac{1}{\sqrt{\det G}}$ has to be
included  into the Hamiltonian density
(\ref{Hdcl}). In other words, the
condition that (\ref{Vacfun}) should be
annihilated by $H$ can be also obeyed
by Hamiltonian in the form $\sim \int
d^p\bx
 \hat{Q}^\dag_M \hat{g}^{MN}\hat{Q}_N$
that however does not lead to
diffeomorphisms invariant theory.

The additional symmetry of the action
 (\ref{S}) is  \emph{global time
translation}
\begin{equation}
t'=t+\delta t \ , \delta
t=\mathrm{const} \nonumber
\\
\end{equation}
as follows from the
fact that \begin{equation}
\Phi'^M(t',\bx)= \Phi^M(t,\bx) \ ,
\quad
\partial_{t'}\Phi'^M(t', \bx)=
\partial_t \Phi^M(t,\bx) \ .
\end{equation}
\section{Foliation-Preserving
Diffeomorphisms} \label{fourth} We
argued in previous section that the
critical $(p+1)$ brane theory is
invariant under \emph{local} spatial
diffeomorphisms and under \emph{global}
time translation. However it turns out
that in order to take into account
appropriately
 the fact that the
$p+1$ dimensional theory is
diffeomorphism invariant we have to
extend these symmetries to  space-time
diffeomorphisms that respect the
preferred codimension-one foliation
$\mathcal{F}$ of world-volume theory by
the slices of fixed time. After this
extension we can develop the
Hamiltonian formalism where the
constraints related to the
diffeomorphisms invariance arise in
natural way.

 By definition
such a foliation-preserving
diffeomorphisms consist  space-time
dependent spatial diffeomorphisms as
well as time-dependent time
reparameterization that are now
generated by infinitesimal
transformations
\begin{equation}\label{fpd}
\delta x^i=x'^i-x^i=\zeta^i(t,\bx) \ ,
\quad  \delta t=t'-t=f(t) \ .
\end{equation}
Note also that the field
$\Phi^M$ is scalar of world-volume
theory and hence
\begin{equation}
\Phi'^M(t',\bx')=\Phi^M(t,\bx)
\end{equation}
so that
\begin{eqnarray}
\partial_{t'}\Phi'(t',\bx')&=&
\partial_t\Phi^M(t,\bx)-\partial_t
\Phi^M(t,\bx)\dot{f}
-\partial_i \Phi^M(t,\bx)\dot{\zeta}^i
\nonumber \\
\partial_{x'^i}\Phi'^M(t',\bx')&=&
\partial_i \Phi^M(t,\bx)-
\partial_j
\Phi^M(t,\bx)\partial_i\zeta^j(t,\bx)
\nonumber \\
\end{eqnarray}
and we see that these objects do
not transform covariantly under
(\ref{fpd}).
Note that under such a diffeomorphism
the element $dt d^p\bx$ transforms as
\begin{eqnarray}
dt'd^p\bx'=
(1+\dot{f})(1+\partial_i \zeta^i)dt
d^p\bx \ .
\nonumber \\
\end{eqnarray}
It can be also easily shown that
$\sqrt{\det G}$ transforms as
\begin{eqnarray}
\sqrt{\det G'(\Phi'(\bx') )}= \sqrt{\det G(\Phi(\bx) )}
(1-
\partial_i\zeta^i(\bx))
\nonumber \\
\end{eqnarray}
and consequently we find that
 \begin{eqnarray}
d^p\bx'\sqrt{\det G'(\Phi'(\bx') )}=
d^p\bx\sqrt{\det G(\Phi(\bx) )} \ .
\nonumber \\
\end{eqnarray}
However due to the fact that the time-derivative
of $\Phi$ transforms non-covariantly under
foliation-preserving diffeomorphism we should
introduce new gauge fields $N_i, N$. It is
convenient to derive their transformation
properties under foliation-preserving diffeomorphism
 from relativistic
 diffeomorphism transformations
of $p+1$ dimensional
metric $g_{\mu\nu}$ by restoring the speed of light $c$
and taking the non-relativistic limit $c\rightarrow \infty$.
This procedure has been nicely reviewed in
\cite{Horava:2008ih} where  the following
transformations rules for $N$ and $N^i$ were
derived
\begin{eqnarray}\label{NNi}
N'(t',\bx')&=&N(t,\bx)(1- \dot{f}) \ , \nonumber \\
 N'^i(t',\bx')
&=& N^i(t,\bx)+N^j(t,\bx)\partial_j
\zeta^i(t,\bx)-
N^i(t,\bx)\dot{f}-\dot{\zeta}^i(t,\bx) \
\nonumber \\
\end{eqnarray}
and consequently
\begin{equation}
dt' N'(t',\bx')=
dt N(t,\bx) \ .
\end{equation}
Further, the form of these
transformations (\ref{NNi})
suggest that it is natural to
 introduce following object
\begin{equation}
\frac{1}{N(t,\bx)}[\partial_t \Phi^M(t,\bx)-N^i(t,\bx)
\partial_i \Phi^M(t,\bx)] \
\end{equation}
that is invariant under folliation preserving
diffeomorphism
\begin{eqnarray}
& &\frac{1}{N'(t',\bx')} [\partial_{t'}
\Phi'^M(t',\bx')-N^{'i}(t',\bx')
\partial_{i'} \Phi^M(t',\bx')]=
\nonumber \\
&=&\frac{1}{N(t,\bx)}[\partial_t
\Phi^M(t,\bx)-N^i(t,\bx)
\partial_i\Phi^M(t,\bx)] \ .  \nonumber \\
\end{eqnarray}
Using these results we can
finally write the  $p+1$ dimensional
action that   is invariant under
folliation-preserving diffeomorphism
\begin{eqnarray}\label{actI}
S&=&\int dt d^p\bx N \sqrt{\det G}\left[
\frac{1}{2\kappa^2}
\frac{1}{N^2}(\partial_t \Phi^M
-N^i\partial_i\Phi^M) g_{MN}
(\partial_t \Phi^N-N^j\partial_j
\Phi^N)-\right. \nonumber \\
&-&\frac{\kappa^2}{8} (\partial_M
G_{ij}G^{ji}-\frac{1}{\sqrt{\det G}}
\partial_i [g_{MK}\partial_j \Phi^K
G^{ji}\sqrt{\det G}])g^{MN}\times
\nonumber \\
& &\left.\times (\partial_N
G_{kl}G^{lk}-\frac{1}{\sqrt{\det G}}
\partial_k[g_{NL}\partial_l \Phi^L
G^{lk}\sqrt{\det G}])\right] \ .
\nonumber \\
\end{eqnarray}
\section{Hamiltonian formalism}\label{fifth}
In this section we develop the Hamiltonian
formulation of the theory that is governed
by the action (\ref{actI}). As the first step we
introduce the momenta $\pi_N,\pi_i$ that
are  conjugate to $N$
and $N^i$ with corresponding
Poisson brackets
\begin{equation}
\pb{N(\bx),\pi_N(\by)}=\delta(\bx-\by)
\ , \quad \pb{N^i(\bx),\pi_j(\by)}=
\delta^i_j\delta(\bx-\by) \ .
\end{equation}
Then, due to the fact that the action
(\ref{actI}) does not contain
 time derivatives of $N$
and $N^i$ we find that $\pi_N$ and
$\pi_i$ are primary constraints of the
theory
\begin{equation}\label{primary}
\pi_N(\bx)\approx 0 \ , \quad
\pi_i(\bx)\approx 0 \ .
\end{equation}
As the next step we determine  the momenta conjugate to
$\Phi^N(\bx)$ from
 (\ref{actI})
\begin{eqnarray}
\Pi_N(\bx)=\frac{1}{\kappa^2N}
\sqrt{\det G}g_{MN}(\partial_\tau
\Phi^N- N^i\partial_i\Phi^N) \ .
\nonumber \\
\end{eqnarray}
Then using the Lagrangian density given
in (\ref{actI}) we find corresponding
Hamiltonian density
\begin{eqnarray}
\mH&=&\partial_t \Phi^N\Pi_N-\mL=
N\left[
\frac{\kappa^2}{2\sqrt{G}}\Pi^Mg_{MN}\Pi^N+\right.
\nonumber \\
&+&\frac{\kappa^2}{8}\sqrt{\det G}
(\partial_M
G_{ij}G^{ji}-\frac{1}{\sqrt{\det G}}
\partial_i [g_{MK}\partial_j \Phi^K
G^{ji}\sqrt{\det G}])g^{MN}\times
\nonumber \\
&\times & \left.(\partial_N
G_{kl}G^{lk}-\frac{1}{\sqrt{\det G}}
\partial_k[g_{NL}\partial_l \Phi^L
G^{lk}\sqrt{\det
G}])\right]+N^i\partial_i \Phi^N\Pi_N \ .
\nonumber \\
\end{eqnarray}
Now the consistency of the primary
 constraints (\ref{primary}) with their time
evolution implies the secondary
constraints:
\begin{eqnarray}
\partial_t \pi_N(\bx)& = &
\pb{\pi_N(\bx),H}=- \left[
\frac{\kappa^2}{2\sqrt{G}}\Pi^Mg_{MN}\Pi^N-
\right.\nonumber \\
&-&\frac{\kappa^2}{8}\sqrt{\det G}
(\partial_M
G_{ij}G^{ji}-\frac{1}{\sqrt{\det G}}
\partial_i [g_{MK}\partial_j \Phi^K
G^{ji}\sqrt{\det G}])g^{MN}\times
\nonumber \\
&\times & \left.(\partial_N
G_{kl}G^{lk}-\frac{1}{\sqrt{\det G}}
\partial_k[g_{NL}\partial_l \Phi^L
G^{jk}\sqrt{\det G}])\right]\equiv
-T_0\approx 0
\nonumber \\
\end{eqnarray}
and
\begin{eqnarray}
\partial_t \pi_i(\bx)=
\pb{\pi_i(\bx),H}= -\partial_i
\Phi^N\Pi_N(\bx)\equiv -T_i\approx 0
\nonumber \\
\end{eqnarray}
However following \cite{Horava:2008ih}
 we suggest another class of
constraints using the fact that $T_0$
can be written as
\begin{eqnarray}\label{T_0Q}
T_0&=&\frac{\kappa^2}{2} Q^\dag_M
\frac{g^{MN}}{\sqrt{\det G}} Q_N \ ,
\quad  Q_M=
i\Pi_M+\frac{1}{2}\frac{\delta
W}{\delta \Phi^M}=
\nonumber \\
&=& i\Pi_M+ \frac{1}{2}\sqrt{\det
G}\left[\partial_M
G_{ij}\Gi^{ji}-\frac{1}{ \sqrt{\det
G}}\partial_i[g_{MN}\partial_j\Phi^N
\Gi^{ji}\sqrt{\det G}]\right] \ .
\nonumber \\
\end{eqnarray}
Then it turns out to be convenient
to solve the
consistency equations for $\pi_N$ and
$\pi_i$ with collections of secondary
constraints
\begin{equation}
Q_M\approx 0 \ , T_i\approx 0 \
\end{equation}
instead of $T_0,T_i$.

Now we would like to demonstrate that
the consistency of time  evolution of
these constraints does not generate any
additional constraints. First of all
it is easy to see that Poisson brackets
between $T_i,Q_M$ and
 $\pi_N, \pi_i$ vanish. Further
we  calculate
the Poisson brackets between $T_i's$ and
we find
\begin{eqnarray}\label{pbTij}
\pb{T_i(\bx),T_j(\by)}
=T_i(\bx)\partial_j\delta(\bx-\by)+
T_j(\bx)\partial_i \delta(\bx-\by)+
\partial_i T_j(\bx)\delta(\bx-\by) \ ,
\nonumber \\
\end{eqnarray}
where we used the basic identities
\begin{eqnarray}
\partial_{y^i}\delta(\bx-\by)&=&-
\partial_{x^i}\delta(\bx-\by) \ ,
\nonumber \\
\partial_{x^i}\delta(\bx-\by)
f(\by)&=&
\partial_{i}\delta (\bx-\by)f(\bx)+
\partial_i f(\bx)\delta(\bx-\by) \ .
\nonumber \\
\end{eqnarray}
Alternatively, we can introduce the
smeared form of the constraints $T_i$
when we introduce the object
\begin{equation}
\mathbf{T}_\zeta =\int d^p\bx
\zeta^i(\bx)T_i(\bx) \ .
\end{equation}
Then the  smeared  form of the
Poisson bracket (\ref{pbTij}) is
\begin{eqnarray}
\pb{\mathbf{T}_\zeta,\mathbf{T}_\eta}=
\int d^p\bx
(\zeta^i\partial_i\eta^k-\eta^i\partial_i\zeta^k)T_k(\bx) \ .
\nonumber \\
\end{eqnarray}
Further, let us consider the Poisson
bracket of $\Pi_M(\bx)$ with any
general functional $F(\Phi(\by))$.
Using the definition of Poisson bracket
\begin{equation}
\pb{\Pi_M(\bx),F(\Phi(\by))}=
-\frac{\delta F(\Phi(\by))}{\delta
\Phi^M(\bx)} \
\end{equation}
and the  fact that
the functional derivative commute
\begin{eqnarray}
\frac{\delta}{\delta \Phi^M(\by)}
\frac{\delta F}{\delta \Phi^N(\bx)}-
\frac{\delta}{\delta \Phi^N(\bx)}
\frac{\delta F}{\delta \Phi^M(\by)}=0 \
 \nonumber \\
\end{eqnarray}
we obtain
\begin{eqnarray}\label{pbQMN}
\pb{Q_M(\bx),Q_N(\by)}&=& -\frac{i}{2}
\frac{\delta^2 W}{\delta
\Phi^M(\bx)\Phi^N(\by)}+
\frac{i}{2}\frac{\delta^2 W}{\delta
\Phi^N(\by)\delta \Phi^M(\bx)}= 0
\nonumber \\
\pb{Q_M(\bx),Q^\dag_N(\by)}&=&-i
\frac{\delta^2 W}{\delta
\Phi^N(\bx)\Phi^M(\by)} \ . \nonumber
\\
\end{eqnarray}
As the next step we determine the
Poisson bracket between $Q_M$ and
$\mathbf{T}_\zeta$.
In order to see the physical meaning of
$\mathbf{T}_\zeta$ let us calculate the Poison
bracket of $\mathbf{T}_\zeta $ with any
function  $F$ of   $\Phi$
\begin{eqnarray}
\pb{\mathbf{T}_\zeta, F(\Phi(\bx))}
= - \zeta^i(\bx)\partial_{i}\Phi^N(\bx)
=-\zeta^i(\bx)\partial_{i} F(\Phi(\bx)) \ .
\nonumber \\
\end{eqnarray}
This result shows that   $\bT_\zeta$ is the generator
of spatial diffeomorphism
$x'^i=x^i+\zeta^i(\bx)$ under which any
scalar function $F(\Phi)$ transforms as
\begin{equation}
\delta_{\mathbf{T}_\zeta}F(\Phi(\bx))=
F(\Phi'(\bx))-F(\Phi(\bx))= -\zeta^i
\partial_i F(\Phi(\bx))=
-\zeta^i\partial_i \Phi^N\frac{\delta
F}{\delta \Phi^N(\bx)} \ .
\end{equation}
We can also study the action of
$\mathbf{T}_\zeta$ on more general
world-volume tensors. For example, the
Poisson bracket of $\mathbf{T}_\zeta$ with
$G_{ij}(\bx)\equiv G_{MN}(\Phi(\bx))
\partial_i\Phi^M(\bx)\partial_j\Phi^N(\bx)$ is equal to
\begin{eqnarray}
& &\pb{\mathbf{T}_\zeta,G_{ij}(\bx)}=
-\zeta^i(\bx)\partial_k\Phi^K(\bx)
\partial_K
G_{MN}(\bx)\partial_k\Phi^M(\bx)
\partial_j\Phi^N(\bx)-
\nonumber \\
&-&G_{MN}(\bx)\partial_i[\zeta^k(\bx)\partial_k\Phi^M(\bx)]
\partial_j\Phi^N(\bx)-
G_{MN}(\bx)\partial_i\Phi^M(\bx)
\partial_j[\zeta^k(\bx)\partial_k\Phi^N(\bx)]=
\nonumber \\
&=&-\zeta^i(\bx)\partial_i G_{ij}(\bx)-
\partial_i\zeta^k(\bx)
G_{kj}(\bx)-G_{ik}(\bx)\partial_j \zeta^k(\bx)\equiv
\delta_{\mathbf{T}_\zeta}G_{ij}(\bx)
\nonumber \\
\end{eqnarray}
that is clearly correct form of the
variation of  $p$ dimensional
metric under diffeomorphism.
Then it is straightforward to calculate
the Poisson bracket between
$\mathbf{T}_\zeta$ and $\Pi_M(\bx)$ and $ \frac{\delta
W}{\delta \Phi^M(\bx)}$ and we find
\begin{eqnarray}\label{TW}
\pb{\mathbf{T}_\zeta
,\frac{\delta W}{\delta \Phi^M(\bx)}}&=&
-\zeta^i(\bx)\partial_i \left[\frac{\delta
W}{\delta \Phi^M(\bx)}\right]- \frac{\delta
W}{\delta \Phi^M(\bx)}\partial_i
\zeta^i(\bx) \ ,
\nonumber \\
\pb{\mathbf{T}_\zeta,\Pi_M(\bx)}&=&
-\zeta^i(\bx)
\partial_i\Pi_M(\bx)-\Pi_M(\bx)
\partial_i\zeta^i(\bx) \
\nonumber \\
\end{eqnarray}
and we finally obtain
\begin{eqnarray}\label{Q,T}
 \pb{\mathbf{T}_\zeta,
Q_M(\bx)}
=-\partial_i Q_M(\bx)\zeta^i(\bx)-
Q_M(\bx)\partial_i\zeta^i(\bx) \ .  \nonumber \\
\end{eqnarray}
Then it is  easy to see that
\begin{eqnarray}
& &\pb{Q_M(\bx),\int d^p\by N T_0(\by)}
=\nonumber \\
&=&\frac{\kappa^2}{2} \int d^p\by N
\pb{Q_M(\bx),Q_P^\dag(\by)}
\frac{g^{PQ}(\by)}{\sqrt{\det
G(\by)}}Q_Q(\by)+\nonumber \\
&+&\frac{\kappa^2}{2} \int d^p\by  N
Q_P^\dag(\by)
\pb{Q_M(\bx),\frac{g^{PQ}(\by)}{\sqrt{\det
G(\by)}}} Q_Q(\by)+
\nonumber \\
&+&\frac{\kappa^2}{2}
\int d^p\by NQ_P^\dag(\by)\frac{g^{PQ}(\by)}
{\sqrt{\det G(\by)}}
\pb{Q_M(\bx),Q_Q(\by)}\approx 0 \ , \nonumber \\
\end{eqnarray}
where the first and the second terms
 vanish on constraint surface
$Q_M\approx 0$ and the third term vanishes
due to the Poisson bracket (\ref{pbQMN}). Then
using (\ref{Q,T}) we find that
$\partial_t Q_M(\bx)=\pb{Q_M(\bx),H}\approx 0$.
In the same way we can show that the Poisson
bracket between $\mathbf{T}_\zeta$ and $H$ vanishes
on constraint surface.  These results imply
that the consistency of time evolutions
of $Q_M$ and $\mathbf{T}_\zeta$ does not
generate additional constraints.

For further purposes we also determine
the Poisson bracket between $\mathbf{T}_\zeta$ and
$W$. Using (\ref{Q,T}) we find
\begin{eqnarray}\label{TWw}
\pb{\mathbf{T}_\zeta,W}&=&
\frac{1}{2}\int d\bx
\pb{\mathbf{T}_\zeta,G_{ij}(\bx)}
G^{ji}(\bx)\sqrt{\det G(\bx)}=
\nonumber \\
&=&-\int d\bx
[\frac{1}{2}\zeta^k(\bx)\partial_k G_{ij}(\bx)
G^{ji}(\bx)\sqrt{\det G(\bx)}
+\partial_i\zeta^i(\bx)
\sqrt{\det G(\bx)}]=
\nonumber \\
&=&-\int d\bx \partial_i[\zeta^i(\bx) \sqrt{\det G(\bx)}]=
0 \ .
\nonumber \\
\end{eqnarray}
In fact this is expected  result since
$W$ is diffeomorphism invariant by construction.

Now we are ready to perform some
preliminary steps in the  quantization of given theory.
Since the Hamiltonian is sum of the  first
class constrains we have to demand that
 each wave functional
of the system should be annihilated by
all these constraints. In fact, we find
previously that the ground state
functional
\begin{equation}
\Psi_0[\Phi^M(\bx)]=\exp[
-\frac{1}{\kappa_W^p}\int d^p\bx
\sqrt{\det G } ]
\end{equation}
satisfies the constraints
\begin{equation}
\hat{Q}_M(\bx)\Psi_0[\Phi(\bx)]=
\left(\frac{\delta}{\Phi^M(\bx)}
+\frac{\delta W}{2\delta \Phi^M(\bx)}\right)\Psi_0[
\Phi(\bx)]=0 \ .
\end{equation}
Further, the operator $\hat{\mathbf{T}}_\zeta$
has following  form in Schr\"{o}dinger
representation
\begin{equation}
\hat{\mathbf{T}_\zeta}=-i\int d\bx \zeta^i(\bx)
\partial_i \Phi^N(\bx)\frac{\delta }{\delta
\Phi^N(\bx)} \ .
\end{equation}
Then it is clear that $\hat{\mathbf{T}}_\zeta$ annihilates
$\Psi_0[\Phi]$ since
\begin{equation}
\hat{\mathbf{T}}\Psi_0[\Phi(\bx)]=i
\int d\by \zeta^i(\by)\partial_i
\Phi^N(\by)\frac{\delta W}{\delta \Phi(\by)}
\Psi_0[\Phi(\bx)]=0
\end{equation}
using the fact that $\int d\bx \zeta^i(\bx)\partial_i
\Phi^N(\bx)\frac{\delta W}{\delta \Phi(\bx)}$
is equivalent to the Poisson bracket
(\ref{TWw})  that vanishes due to the diffeomorphism
invariance of $W$.
In other words the vacuum wave functional obeys
all constraints of the theory. Further, since it
is annihilated by $\hat{Q}_M$ it is also eigenstate
of the Hamiltonian with zero energy. On the other
hand it is an open problem whether this is a normalizable
state and whether  there are more general
functionals that have non-zero energy with respect
to given Hamiltonian.
\section{Conclusion}\label{sixth}
This paper is devoted to the
construction of new class of $p+1$
dimensional non-relativistic theories
that obey the detailed balance
condition that claims that their
potential is derived from the variation
principle of $p$ dimensional Nambu-Goto
form of p-brane action. We also
extended symmetries of given action and
construct $p+1$ dimensional action that
is invariant under foliation-preserving
diffeomorphisms.

We hope that these new non-relativistic theories
have many interesting properties and should be studied
further. In particular, it will be interesting
to study  their
quantum properties in more details.
We would also like to extend
this formalism to the case of
 BPS and non-BPS Dp-branes and to the
case of topological p-branes, following
for example \cite{Bonelli:2005rw}.
We also mean that
it would be  interesting to study the dynamics of
these non-relativistic
p-branes in backgrounds with the metric that
does not have Euclidean signature. We hope to return
to these problems in future.
\vskip .2in \noindent {\bf
Acknowledgements:} This work was
 supported by the Czech
Ministry of Education under Contract
No. MSM 0021622409. I would also like
to thank the Max Planck Institute at
Golm for its kind hospitality during
finishing of this work.

\newpage

\end{document}